\documentstyle[12pt]{article}
\def\fun#1#2{\lower3.6pt\vbox{\baselineskip0pt\lineskip.9pt
\ialign{$\mathsurround=0pt#1\hfil##\hfil$\crcr#2\crcr\sim\crcr}}}

\newcommand{\cen}{\centerline}
\newcommand{\bc}{\begin{center}}
\newcommand{\ec}{\end{center}}
\newcommand{\bd}{\begin{displaymath}}
\newcommand{\ed}{\end{displaymath}}
\newcommand{\be}{\begin{equation}}
\newcommand{\ee}{\end{equation}}
\newcommand{\ba}{\begin{array}}
\newcommand{\ea}{\end{array}}
\newcommand{\bt}{\begin{tabular}}
\newcommand{\et}{\end{tabular}}
\newcommand{\un}{\underline}

\newcommand{\e}{\mbox{e}}

\begin{document}
~
\vspace{-0.5cm}
\begin{flushright}
\begin{large}
ITEP-TH-41/96\\
\end{large}
\end{flushright}
\vspace{0.5cm}
\bc
{\large\bf BREAKDOWN OF ZERO--MODE APPROXIMATION\\[0.2cm]
IN THE INSTANTON VACUUM}\\[0.4cm]
{\it B.O.Kerbikov, D.S.Kuzmenko, Yu.A.Simonov\\
Institute of Theoretical and Experimental Physics,\\
117259 Moscow, Russia}
\vspace{0.8cm}

\parbox{14cm}{~~~~~A new approach to effective theory of quarks in the
instanton vacuum is presented. Exact equations for the quark
propagator and Lagrangian are derived which contain contributions of
all quark modes with known coefficients. The resulting effective
Lagrangian resembles that of the Nambu--Jona--Lasinio model. The
commonly used zero--mode approximation is shown to be invalid in the
chiral limit.}

\ec
\vspace{0.8cm}


1. The recent lattice data [1] provided evidence that instantons may
be responsible for nonperturbative behavior of $q\bar q$ correlators
[2] which makes the study of the quark dynamics in the instantonic
vacuum [3--5] an important and fundamental problem.

To date practically all papers on the subject have relied upon the
use of the so--called zero--mode approximation (ZMA) which amounts to
comprising only zero quark mode in a single--instanton fermion
propagator [3]. Correspondingly an ansatz for the partition function
and effective quark Lagrangian (EQL) containing zero modes only have
been proposed [5] and widely used in literature [6,7].

The purpose of this letter is to give a complete normal mode
expansion of the EQL and of the quark propagator. For a specific
choice of the coefficients in this expansion one retrieves the ansatz
[5] for EQL. However this option is invalidated by exact calculation
presented below which shows that zero--mode term vanishes in the
chiral limit and hence ZMA for EQL breaks down. More intricate is the
analysis of the quark propagator $S$ in the
instanton--anti--instanton vacuum. Here the  zero--mode term survives
but enters with a coefficient depending solely on contributions of
higher modes. The same feature can be seen in the quark partition
function $<det S^{-1}>$, where the average $<\ldots>$ is defined
below. Thus ZMA breaks down and the new quark dynamics associated
with nonzero modes emerges. The main features of this dynamics are
outlined in what follows.

2. To make discussion transparent consider an ideal instanton gas
with the superposition ansatz [8--10] and with equal number of
instantons and anti--instantons, $N_+=N_-=N/2$:

\be
A_{\mu}(x)=\sum^N_{i=1}A^{(i)}_{\mu}(x-R_i),
\ee

\be
gA^{(i)}_{\mu}=\frac{\bar\eta_{a\mu\nu}(x-R_i)_{\nu}\rho^2\Omega^+_i
\tau_a\Omega_i}{(x-R_i)^2[(x-R_i)^2+\rho^2]},
\ee
where $\Omega_i$, $R_i$, and $\rho$ are color orientation, position
and scale--size of the $i$--th instanton.

The EQL is obtained from the Euclidean partition function after
averaging over $\{\Omega_i,R_i\}$:

\be
Z=\int D\psi D\psi^+\e^{-\int dx\psi^+S^{-1}\psi}\prod^N_{i=1}
\frac{dR_i}{V}d\Omega_i=\int D\psi D\psi^+\e^{-L_{eff}},
\ee
where definitions here and in what follows are:

\be
\ba{l}
S^{-1}_0=(-i\hat\partial-im_f),\\[0.3cm]
S^{-1}_i=(-i\hat\partial-g\hat A^{(i)}-im_f),\\[0.3cm]
S^{-1}=(-i\hat\partial-g\hat A-im_f).
\ea
\ee
Next we introduce the standard set of eigenfunctions $\{u^i_n\}$,
$n=0,1,2,\ldots$

\be
(-i\hat\partial-g\hat A^{(i)})|u^i_n>=\lambda_n|u^i_n>,
\ee
and expand $S^{-1}$ given by (4) as

\be
S^{-1}=S^{-1}_0+\sum_{i,m,n}S^{-1}_0|u^i_m>\varepsilon^i_{mn}
<u^i_n|S^{-1}_0,
\ee
where $\hat\varepsilon$ can be represented either as

\be
\varepsilon^i_{mn}=<u^i_m|S_0S^{-1}_iS_0-S_0|u^i_n>=
-<u^i_m|(S_i-S_0)[1+S^{-1}_0(S_i-S_0)]^{-1}|u^i_n>,
\ee
or simply as

\be
\varepsilon^i_{mn}=-g<u^i_m|S_0\hat A^{(i)}S_0|u^i_n>.
\ee
Performing in (3) averaging with the help of cumulant or cluster
expansion, one obtains $L_{eff}$ in the form

\be
L_{eff}=-\int dx\psi^+S^{-1}_0\psi+\sum^{\infty}_{n=2}(-1)^n2^n
\left(\frac VN\right)^{n-1}\sum_{fmm'}\int d\Gamma_n
det^{(n)}_{k,l}J_{kl},
\ee
where

\be
d\Gamma_n=\prod^n_{j=1}\frac{dp_j}{(2\pi)^4}\cdot\frac{dp'_j}{(2\pi)^4}
(2\pi)^4\delta\left(\sum_j(p_j-p'_j)\right),
\ee

\be
J_{kl}=\left(\psi^{f_k}(p_k)\right)^+M^{f_kf_l}_{mm'}(p_k,p'_l)
\psi^{f_l}(p'_l),
\ee
and similarly to [5,6] we have introduced the vertices

\be
M^{gr}_{mm'}(p,p')=\frac{N}{2VN_c}(\hat
p-im_g)\varphi_m(p)\varepsilon^i_{mm'}\varphi^+_{m'}(p')(\hat
p'-im_r),
\ee
with $\varphi_m(p)$ being the form factor of $u^i_m$ in momentum
space.

Summation in (9) starts from $n=2$ since the $n=1$ term drops out as
a result of integration over color orientations.

The EQL in (9) is a sum of $n\times n$ determinants. If one confines
oneself to ZMA i.e. puts $\varepsilon^i_{00}$ finite and
$\varepsilon^i_{m>0,n>0}$ equal to zero the sum runs only over $n\le
N_f$. This restriction is due to Grassmann nature of $J_{kl}$. Thus
even in ZMA one obtains e.g. for $N_f=3$ three $2\times2$
determinants and one $3\times3$ determinant. Only the last one is
present in the ansatz [5] with the identification
$\varepsilon^i_{00}\equiv\varepsilon$, $M_{00}\equiv M$. Therefore
our results are in contrast to the common lore according to which for
a given number of flavors $N_f$ the only vertex appearing in the
chiral limit contains $2N_f$ quark operators. We can reproduce this
result for $N_f=2$ if only $\varepsilon^i_{00}$ is kept nonzero,
while for $N_f=3$ this conjecture does not suffice and we get
additional $4q$ terms.

Now we come to the central point of the letter. Let us estimate
$\varepsilon^i_{00}$ using (8). In the chiral limit the term $S_0\hat
A^{(i)}S_0$ is chirally odd and therefore $\varepsilon^i_{00}$
vanishes, for $m_f\ne0$ one has

\be
\varepsilon^i_{00}=O(m_f),\qquad m_f\to0.
\ee
At the same time nonzero modes $u^i_{mn}$ do not have definite
chirality and hence matrix elements $\varepsilon^i_{mn}$ do not
vanish as $m_f\to0$. Thus all terms proportional to
$\varepsilon^i_{00}$ in $L_{eff}$ vanish in the chiral limit implying
the breakdown of ZMA in the effective quark Lagrangian.

3. Now we turn to the quark propagator, expressing it again through
$\varepsilon^i_{mn}$. Inverting (6) one finds

\be
S=S_0-\sum_{ijmn}|u^i_m>\left(\frac{1}{\hat\varepsilon^{-1}+\hat
V}\right)^{ij}_{mn}<u^j_n|,
\ee
where $(\hat\varepsilon)^{ij}_{mn}=\delta_{ij}\varepsilon^i_{mn}$,
and

\be
(\hat V)^{ij}_{mn}=<u^i_m|S^{-1}_0|u^j_n>.
\ee
Note that summation in (14) extends over different instantons and
hence over $u^i_0$ and $u^j_0$ of different chiralities. Equation
(14) has to be compared to the following expression common to most
papers on the subject [3,5,6]

\be
S=S_0-\sum_{i,j}|u^i_0>\left(\frac{1}{2im+V}\right)^{ij}_{00}<u^j_0|,
\ee
which contains only zero modes contributions. To derive (16) one
starts with the following approximation for the quark propagator in a
single instanton field [3,5]:

\be
S_i=(-i\hat\partial)^{-1}+\frac{|u^i_0><u^i_0|}{-im}.
\ee
Introducing this ansatz into expression (7) for $\varepsilon^i_{mn}$
we get

\be
\varepsilon^i_{00}=\frac{1}{2im},\qquad
\varepsilon^i_{m>0,n>0}=0.
\ee
Using this form of $\hat\varepsilon$ in (14) one recovers the
standard ZMA (16). Now, comparing (18) to (13) we conclude that
ansatz (17) is unjustified. Actually when
$\varepsilon^i_{00}$ vanishes in the chiral limit in line with (13),
the propagator (14) still contains terms $|u^i_0><u^j_0|$, but with
coefficients depending upon higher modes contributions $V^{ij}_{mn}$.
This can be seen expanding (14) in series in powers of $\varepsilon$,
i.e.

\be
S=S_0-\sum_{i,j,m,n}|u^i_m>(\hat\varepsilon-
\hat\varepsilon\hat V\hat\varepsilon+
\hat\varepsilon\hat V\hat\varepsilon\hat V\hat\varepsilon-
\ldots)^{ij}_{mn}<u^j_n|.
\ee
If one neglects nonzero modes in $V_{mn}$ in (21), then the
coefficient of $|u^i_0><u^j_0|$ automatically vanishes. Thus the
popular ZMA for the propagator (16), using only zero modes for
different instantons and anti--instantons, looses its  ground--this
contribution vanishes in the chiral limit while nonzero modes give  a
nonvanishing result.

It is worth noting that the  consistency of the approximation (17) was
questioned in [9] in connection to the calculation of the two--point
correlation function. It was shown in [9] that it is absolutely
necessary to keep the order $\sim m$ terms in $S_i$. However since
for massive fermions the single  instanton propagator $S_i$ is not
explicitly known the effects of higher modes and finite mass have
been investigated only numerically [11].

Finally let us see the effect of nonzero modes in the quark partition
function, which is obtained from (3) integrating first over quark fields.
Using (6) for $S^{-1}$  one easily  obtains

\be
Z/Z_0=\prod^{N_f}_{f=1}det(1+\hat\varepsilon\hat V),
\ee
where $\hat\varepsilon$ and $\hat V$ are the same matrices as in (14), (15).
It is clearly seen in (20) that the contribution of only zero modes on
different instantons, i.e. $V^{ij}_{00}$ is accompanied by factors
$\varepsilon^k_{00}$, and vanishes in the chiral limit due to (13). Thus
in this limit nonzero modes are necessary to ensure nonvanishing result.

4. One may wonder, why ZMA is invalid even though phenomenologically it
looks like giving reasonable results [5-6, 12]. One of the reasons might be
that $\varepsilon^i_{00}\equiv\varepsilon$ has been treated as a parameter
connected to the properties of the instanton vacuum via the relation
$\varepsilon\sim(\frac{N_cV}{N\rho^2})^{1/2}$ , while the properties
of the vacuum have been in turn adjusted to the correct value of the
gluon condensate.

Our results are at first sight in contradiction to the Banks--Casher relation
[13] which connects the chiral condensate with the density of global (quasi)
zero modes. The standard picture suggests that the later originate from
individual zero modes, and hence would disappear as soon as (13) holds.
However here the standard picture fails. An insight into its failure is
provided by quantum mechanics of collective levels in $N$ potential wells in
$4d$. If each of the well has one loosely bound level and continuum
(equivalent to zero mode and  nonzero modes), then the approximation
of keeping only the bound state poles in the Green's functions of
each well is known to give an  inadequate description of collective
bound states [14].  More than that, the pole approximation is a poor
one even for the Green's function of the individual well, and instead
the so--called unitary pole approximation (UPA) has to be used [15].

5. To summarize , we have outlined the new approach to the effective
theory of quarks in the instanton vacuum. Our EQL is similar to that
of NJL [16], namely it starts from $4q$ term which might play an
important role in phenomenology. Analogy to NJL model calls for
construction of gap equation yielding chiral quark  mass and quark
condensate. Finally, the bosonization procedure has to be performed
yielding the effective chiral Lagrangian for Nambu--Goldstone modes.
This program is in progress now and will be reported elsewhere.

The authors are thankful to A.V.Smilga, Yu.M.Makeenko and S.V.Bashinsky for
useful discussions. The work was supported by INTAS Grant 94--2851
and by the Russian Fund for Fundamental Research Grant 96--02--19184a.\\

\cen{\bf\un{\hspace{5cm}}}
\vspace{0.3cm}

\noindent
\begin{enumerate}
\item M.C.Chu, J.M.Grandy, S.Huang, J.W.Negele, Phys. Rev.
Lett. {\bf70}, 225 (1993); Phys. Rev. {\bf D49}, 6039 (1994).

\item E.V.Shuryak, Rev. Mod. Phys. {\bf65}, 1 (1993).

\item D.Diakonov and V.Petrov, Nucl. Phys. {\bf B272}, 457
(1986).

\item E.V.Shuryak, Nucl. Phys. {\bf B302}, 559, 574, 599
(1988).

\item D.Diakonov and V.Petrov, Preprint LNPI--1153 (1986);\\
D.Diakonov and V.Petrov, in: Quark Cluster Dynamics, Lecture Notes in
Physics, Springer--Verlag (1992) p.288.

\item M.A.Nowak, J.J.M.Verbaarschot, and I.Zahed, Phys. Lett.
{\bf B228}, 251 (1989); Nucl. Phys. {\bf B324}, 1 (1989).

\item Yu.A.Simonov, Yad. Fiz. {\bf57}, 1491 (1994).

\item C.Callan, R.Dashen, and D.Gross, Phys. Rev. {\bf D17},
2717 (1978).

\item N.Andrei and D.J.Gross, Phys. Rev. {\bf D18}, 468 (1978).

\item H.Levin and L.G.Yaffe, Phys. Rev. {\bf D19}, 1225
(1979).

\item E.V.Shuryak and J.J.M.Verbaarshot, Nucl. Phys. {\bf
B410}, 37 (1993).

\item T.Wettig, A.Sch\"afer, and H.A.Weidenm\"uller, Phys.
Lett. {\bf B367}, 28 (1996).

\item T.Banks and A.Casher, Nucl. Phys. {\bf B169}, 103
(1980).

\item A.I.Baz', Ya.B.Zeldovich, and A.M.Perelomov,
{\it Scattering, reactions and decays in nonrelativistic quantum
mechanics}, Moscow, Nauka, 1971;\\ B.Sakita, {\it Quantum theory of
many--variable systems and fields}, World Scientific, 1985.

\item C.Lovelace, Phys. Rev. {\bf135B}, 122 (1964).

\item Y.Nambu and G.Jona--Lasinio, Phys. Rev. {\bf122}, 345
(1961).

\end{enumerate}

\end{document}